\documentclass[review]{elsarticle}

\usepackage{lineno,hyperref}
\modulolinenumbers[5]

\journal{Euro Phys Letters}

\usepackage{amsmath,amssymb}

\newcommand{\fig}[1]{Fig.~\ref{fig:#1}}

\begin{document}

\begin{frontmatter}

\title{Increased adaptability to rapid environmental change can more than make up for the two-fold cost of males}

\author{Caroline M.~Holmes\fnref{email,curraddress}}
\fntext[email]{cholmes@princeton.edu}
\fntext[curraddress]{Current address: Department of Physics, Princeton University, Princeton, NJ}
\author{Ilya Nemenman}
\author{Daniel B.~Weissman\fnref{email1}}
\fntext[email1]{daniel.weissman@emory.edu}
\address{Departments of Physics and Biology,\\ Initiative in Theory and Modeling of Living Systems,\\ Emory University, Atlanta, GA 30322, USA}

\begin{abstract}
The famous ``two-fold cost of sex'' is really the cost of anisogamy -- why should females mate with males who do not contribute resources to offspring, rather than isogamous partners who contribute equally? In typical anisogamous populations, a single very fit male can have an enormous number of offspring, far larger than is possible for any female or isogamous individual. If the sexual selection on males aligns with the natural selection on females, anisogamy thus allows much more rapid adaptation via super-successful males. We show via simulations that this effect can be sufficient to overcome the two-fold cost and maintain anisogamy against isogamy in populations adapting to environmental change. The key quantity is the variance in male fitness -- if this exceeds what is possible in an isogamous population, anisogamous populations can win out in direct competition by adapting faster.
\end{abstract}

\end{frontmatter}

% \linenumbers

\section{Introduction}

In most sexually reproducing species, 
the different sexes contribute different amounts of resources to offspring \cite{lehtonen2016isogamous}.
One fundamental way that they do this is via \textit{anisogamy}:
producing gametes of different sizes, so that one sex contributes more resources to the zygote. 
This anisogamy is in fact what defines the sexes, with females typically defined as the
sex that invests more resources in its offspring \cite{lehtonen2016isogamous}.
Many sexual species take this asymmetry much further,
with females providing essentially all the resources for offspring and
males providing virtually nothing besides half of the offspring's genome.
In this case, assuming that male and female offspring require equal resources to reproduce,
males impose their famous ``two-fold cost'' on females
-- parthenogenetic females could pass on twice as much of their genetic material
as those who mate with males, since they would have the same number of offspring while being responsible for all of their genetic material.

While parthogenetic lineages have a short-term advantage, they
essentially lose the ability to recombine, which can be crucial
for generating variability that can be selected over longer time scales
\cite{weismann1889,fisher1930genetical,muller1932,kimura1966mutational,williams1975sex,MaynardSmith1978,kondrashov1982selection,michod1988evolution,hamilton1990sexual,barton2009}.
However, this fact by itself does not explain the prevalence of males,
as \textit{isogamous} species (or, more generally, those in which
both mating types invest equally in offspring, as in, for example, yeast \citep{greig2009natural}) retain all the benefits 
of recombination while still potentially producing twice as many offspring 
as anisogamous ones \citep{kodricbrown1987}.
Why then are almost no multicellular sexual species isogamous,
given that the fitness cost of anisogamy is so high?

The primary class of explanations for the prevalence of anisogamy is based on
direct selection on the size of gametes and zygotes  \citep{parker,scudo1967adaptive,charlesworth,MaynardSmith1978,matsuda,bell1978evolution,cox1985gamete,lehtonen2016isogamous}.
For example, for widely-separated plants to produce large seeds, 
it will generally be much easier for a pollen grain to travel from one plant
to another than for a half-seed (or for two half-seeds to meet somewhere in the middle). 
Related arguments also provide reasons why there are often exactly two mating types \cite{hoekstra1980organisms}.

Another class of explanations for anisogamy, which we will focus on here,
considers the influence of anisogamy on evolution.
The key process at work is \textit{sexual selection}, which we will use to
refer to the fitness components relating to mating success.
Sexual selection can also exist in isogamous species -- for example,
yeast produce and follow pheromone gradients to find mates \cite{beekman2016sexual}. 
As far as sexual selection acts only to produce assortative mating,
in which high-fitness individuals preferentially mate with other 
high-fitness individuals, there is no necessary advantage for anisogamy,
as isogamous populations can do this as well.
However, because males in anisogamous populations can produce large numbers of sperm with little cost, 
they can experience an additional form of sexual selection, in which some males reproduce many times while others do not 
reproduce at all. 
While this is also possible to a limited extent in isogamous populations, having individuals that do not reproduce necessarily reduces
the resources available for the next generation. 
By contrast, under anisogamy, if the males favored by sexual selection are the ones carrying ``good genes''
that increase other fitness components, this form of sexual selection can 
greatly enhance natural selection without a reduction in the reproductive 
output of the population \citep{fisher1930genetical,zahavi1975,hamilton1982}, 
although it is unclear how often it actually does so in nature (see, e.g., \cite{candolin2008}).

``Good genes'' sexual selection allows anisogamous populations to greatly reduce their
mutational load and the probability that deleterious mutations fix,
potentially overcoming the two-fold cost if deleterious mutation rates are large \cite{manning1984males,whitlock2000fixation,agrawal2001,siller2001sexual,hadany2007,whitlock2009,roze2012,kleiman2015}. 
In other words, males can act as dead ends, where deleterious mutations go to die. 
Sexual selection can also help anisogamous populations adapt by increasing the fixation probability of beneficial mutations \cite{whitlock2000fixation}, but the magnitude of this
effect is likely to be limited, as interference among mutations
generally prevents their rate of incorporation from reaching levels
that would balance the two-fold cost \cite{weissman2012}.
However, this interference limit does not apply to non-equilibrium selection
on standing variation; in this case, even if all beneficial alleles
start at frequencies such that they are essentially certain to be fixed,
sexual selection can provide an advantage by allowing them to be fixed
more rapidly \cite{kodricbrown1987,candolin2008,hollis2009}. 
The importance of this effect has long been known in animal breeders,
who generally use extreme selection on males to rapidly improve stocks, with
the thoroughbred stud Storm Cat, for instance, fathering over 1000 foals \cite{stormcat}.
However, only one previous study has quantitatively considered how
it might provide an advantage for anisogamy:
Lorch et al.~\citep{lorch2003condition}
observed that in simulated populations, sexual selection could
produce a spike in the rate of adaptation to environmental change,
although the observed advantage was not large enough to balance the
two-fold cost.
Here we use simulations of direct competition between isogamous and anisogamous populations to show that the increase in the rate
of adaptation to environmental change can, in fact, be large enough to
single-handedly balance the two-fold cost of males, and we quantify
the conditions for it do so under a minimal model of sexual selection.

\section{Model}

We investigate the fitness effects of anisogamy by numerically simulating competition 
between a sexual anisogamous population and an equivalent sexual but isogamous 
population adapting to a sudden environmental change.
We expect that the isogamous population will initially outcompete
the anisogamous one because of the two-fold cost of males.
However, we also expect that sexual selection will allow the anisogamous population to adapt faster, as the fittest males will produce very large numbers of offspring.
If this increase in speed is large enough, 
the anisogamous population will overcome the two-fold cost before it goes extinct. 
We use simulations and approximate calculations to determine the parameter values
for which this happens.

In our model, the isogamous population has two mating types;
although they invest equally in
offspring, we will refer to them as ``females'' and ``males'' since they play
slightly different roles in the simulations.
The total population size of all individuals together is fixed at $N$
diploid individuals. 
Each generation, each anisogamous female produces $n$ eggs, while 
each isogamous individual produces $2n$ gametes, corresponding to the 
two-fold cost of males. We emphasize that this model accounts for the general asymmetry in the parental investment in offspring, and not just in the gamete size.
Each anisogamous male produces an effectively infinite number of sperm.
In the first stage of selection, females (including isogamous ``females'')
compete with each other to place their gametes in the next generation, with
each gamete being selected with probability proportional to the fitness of its mother.
In the second stage of selection, males compete with each other to fertilize the successful female gametes.
Anisogamous males only fertilize anisogamous female eggs and isogamous
``males'' only fertilize anisogamous ``female'' gametes, so there is no competition
between mating systems in this stage and no interbreeding between the
mating systems. That is, the competition between the isogamous and the anisogamous populations only occurs when selecting the female gametes that will mate and contribute to the following generation.
Male gametes are selected with probability proportional to the fitness of their father.
To keep the sex ratio fixed, each mating produces exactly two offspring, which are genetically identical 
except that one is male and one is female. We do not expect this constraint to significantly affect our conclusions. 

Each individual is diploid, with a genome consisting of $L$ loci.
All loci are unlinked, i.e., gametes sample one of the two parental alleles
at each locus independently.
Each locus is binary, with allele 1 conferring an advantage in log fitness $s$ over allele 0, with no epistasis.
Thus if an individual has genotype $\mathbf X$, with $X_k \in \{0, 1, 2\}$
being the number of 1 alleles that the individual has at locus $k$,
its fitness $w$ is:
\begin{equation}\label{fitness}
	w \equiv \exp\left(s \sum_{k=1}^L X_k\right).
\end{equation}
Technically, this is the individual's breeding value for fitness rather than 
fitness itself (defined as the expected number of offspring), which
also depends on sex and mating system.
Explicitly, while an isogamous individual cannot have more than $2n$ offspring and an anisogamous female cannot more than $n$ offspring, regardless of their value of $w$,
an anisogamous male with extremely high $w$ could, in principle, sire
all of the anisogamous offspring in the next generation.

We are modeling a situation in which an environmental shift has just 
changed the selection on standing variation.
For simplicity, we assume that all the alleles were previously neutral,
with starting frequencies $F_k(t=0)$ drawn independently from the distribution
(\cite{crow1970introduction}, Eq.~(9.3.3)):
\begin{equation}\label{standingvar}
p(f) \propto \left(f (1-f) \right)^{4 N \mu - 1},
\end{equation}
where $\mu$ is the mutation rate per locus per generation.
At the beginning of the simulation, each individual's alleles are drawn
independently at each locus according to their frequency $F_k(0)$, i.e.,
the population is in linkage equilibrium, up to stochastic effects.

\subsection{Simulated parameter values}

For the genome length $L$, we considered between $10$ and $1000$ loci.
The selective advantage $s$ of each allele ranged between 0 and 0.5,
with each simulation run having a single value for all loci.
The loci and alleles in these simulations should be understood as linkage
blocks -- the longest stretches of genome that can be treated as 
effectively unlinked -- rather than individual nucleotides.
For instance, $L=100$ and $s=0.1$ might correspond to a human-sized genome of 3 gigabases, viewed as being composed 
of ``loci'' of 30 megabases each, each potentially contributing variance in log 
fitness of $\mathcal{O}(s^2) \sim 0.01$.
The population size was $N \in [100, 10000]$.
Each simulation started with 
an equal number $N/2$ of anisogamous and isogamous adults and continued
until one mating system drove the other to extinction.
Since the effects described here are largely consistent across population sizes varying over orders of magnitude, as we have verified numerically, for concreteness, 
we use $N=1000$ for all figures. 
We used mutation rate $\mu = 0.01$ in \eqref{standingvar},
but did not actually include new mutations in the course
of our simulations, as their effect is expected to be negligible 
given the large amount of standing variation and the short time
needed for one population to out compete the other -- e.g., $<10$ generations in \fig{example_1run}.
Like $s$, $\mu$ should be understood as an effective parameter describing a linkage 
block rather than an individual nucleotide. 
As long as $N \mu \gtrsim 1$ (as it always is in our simulations),
each linkage block will begin with standing variation and will harbor variance
in log fitness on the order of the maximum value of $\sim s^2$.
As we are concerned with the effect of anisogamy on the maximum possible rate 
of adaptation, the simulated parameter values correspond to extremely strong selection, much stronger than is typically observed in natural populations;
we consider the relationship with natural dynamics in the Discussion.
For every plotted parameter combination, we ran 100 independent simulations to calculate averages, which ensured that statistical fluctuations are much smaller than the means. 

\section{Results}

\begin{figure}
  \centering
  \includegraphics[width=6in]{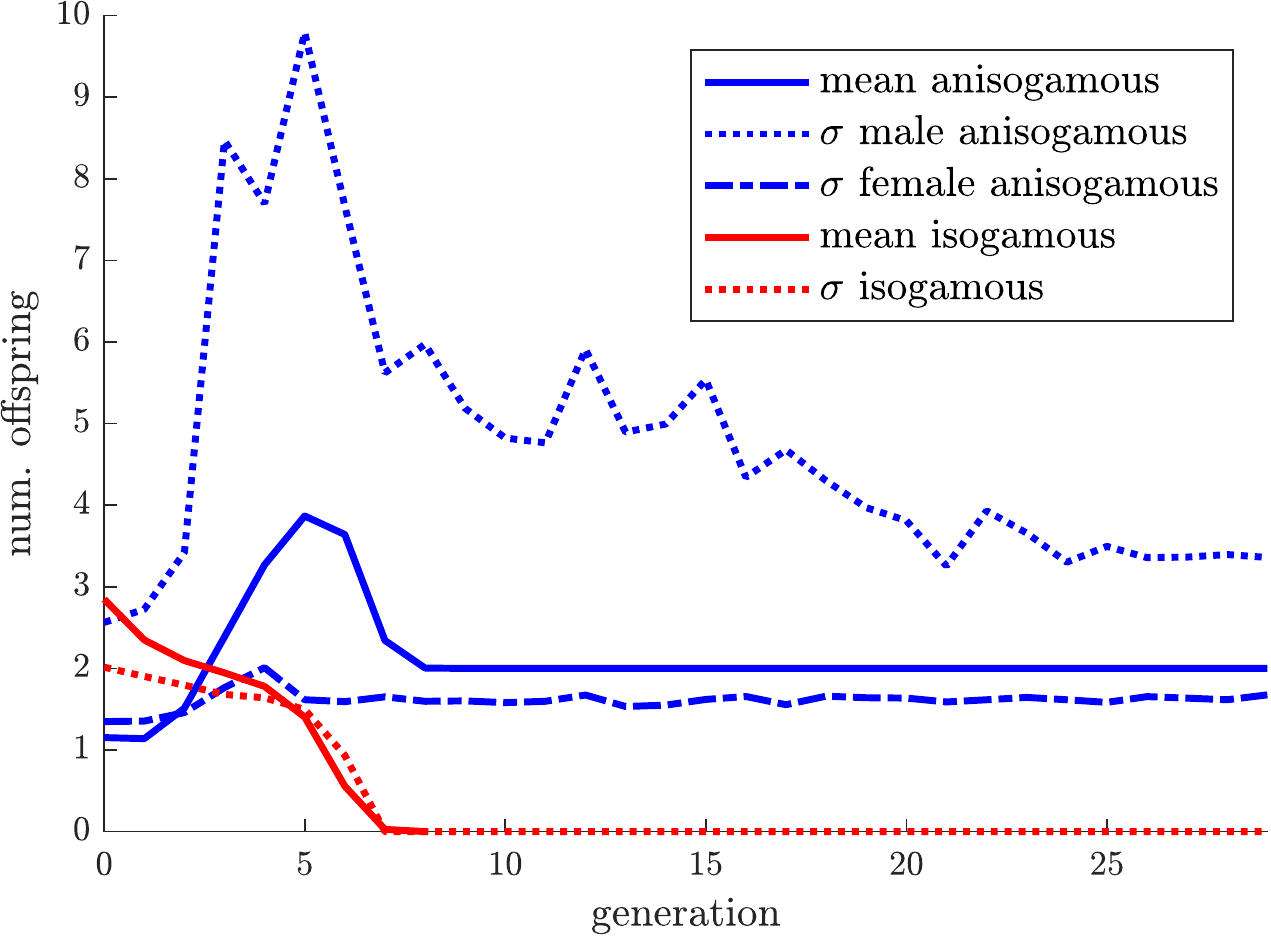}
  \caption{Trajectory of a single typical simulation run, 
  showing the mean and standard deviation of the number of offspring per individual over time for isogamous and anisogamous individuals.
In the first generation, isogamous individuals produce more than twice as many offspring as isogamous individuals due to the two-fold cost of males and
stochastic effects,
but the isogamous population adapts faster and drives them extinct by generation 7. The increased rate of adaptation is driven by the large spike in the variance in 
offspring number among males around generation 5, which cannot be matched by the 
isogamous population because all individuals have upper limits to their reproductive
capacity. Parameters: $N = 1000$, $n = 3$ eggs per anisogamous female, $L = 300$ loci, and $s = 0.3$.}
  \label{fig:example_1run}
\end{figure}

\fig{example_1run} demonstrates dynamics of the realized fitness (number of offspring) of mating systems in a single typical run of the simulation, where anisogamy outcompetes isogamy.
The mean fitness of the isogamous individuals starts at more than twice that of the anisogamous ones, because of the two-fold cost of males and stochasticity in the distribution of alleles and in reproduction (note that the mean fitnesses of anisogamous males and anisogamous females are the same since each offspring has one father and one mother).
However, the mean anisogamous fitness then increases rapidly, eventually winning the competition before settling down at 2, the replacement rate.
The figure also shows that the spike in the mean realized fitness of the anisogamous individuals coincides with a dramatic spike in the standard deviation of the realized fitness distribution of males: with a mean of $\sim 3$ offspring, the standard deviation of offspring per male reaches 10, so that some males can have 20 or more offspring, while others have zero.
At the same time, as expected, the standard deviation of the realized fitness distribution of females barely increases (and always stays significantly below that of the males).
This illustrates that the realized fitness of females is limited since the maximum offspring number is not more than the reproductive capacity $n=3$, and there is, therefore, room to procreate even for not very fit females. 
In addition, this verifies our intuition that it is the variance in the number of offspring for males that drives the fitness increase. 

\begin{figure}
  \centering
  \includegraphics[width=\textwidth]{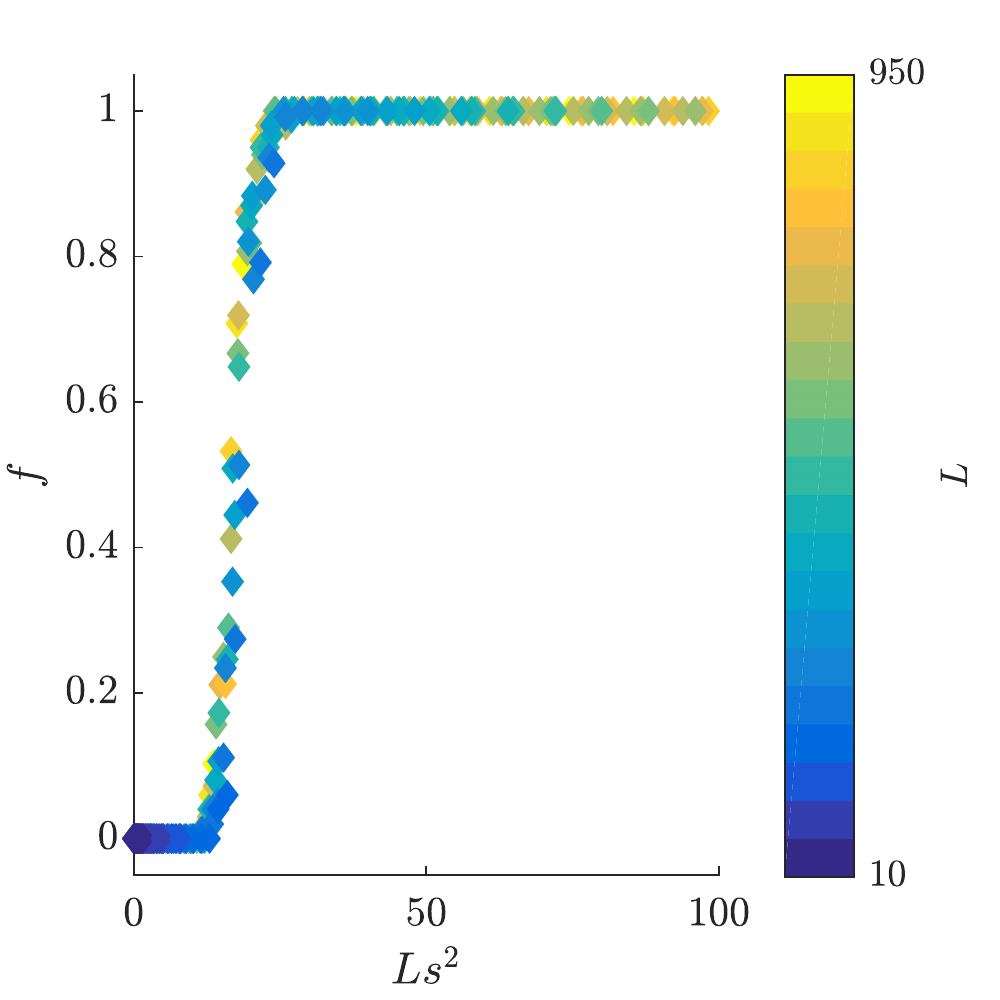}
  \caption{Plot of the frequency of an anisogamous population winning in a competition with the isogamous one over $100$ runs, varying $L$, and $s$. Error bars are not shown, but are the binomial error. The color of the markers represents the associated value of $L$.  This confirms that the parameter $Ls^2$ controls the outcome of the competition between the anisogamous and isogamous populations. Parameters: $N = 1000$, $n = 3$. }
  \label{fig:ls_sqr}
\end{figure}

\begin{figure}
  \centering
  \includegraphics[width=\textwidth]{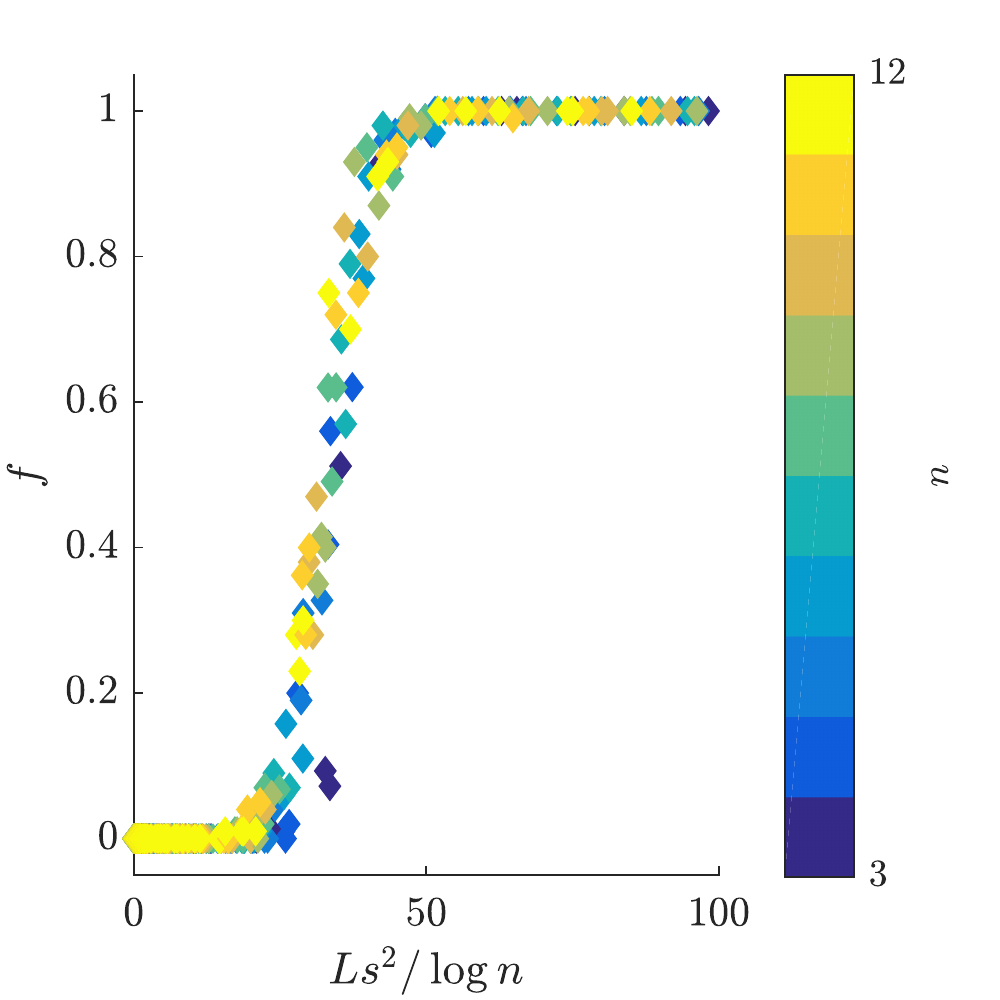}
  \caption{Plot of the frequency of an anisogamous population winning in a competition with the isogamous one over $100$ runs, varying $n$, $L$, and $s$. Like in the previous figure, error bars are not shown, but are the binomial error. The color of the markers represents the associated value of $n$.  This confirms that the compound parameter $Ls^2/\log n$ and not just $Ls^2$ controls the outcome of the anisogamous/isogamous competition. Here, $N = 1000$.}
  \label{fig:ls_sqr_n}
\end{figure}

Fisher's fundamental theorem of natural selection states that the rate of increase of mean fitness is equal to the variance in fitness.
Thus the rate of adaptation is limited to $\mathcal{O}(n^2)$ under isogamy,
while anisogamous populations can adapt faster via selection on males,
and we predict that this difference will affect the success of the anisogamous population once the variance in breeding values
for fitness $w$ approaches $\mathcal{O}(n^2)$, or, equivalently,
$\log{\rm var}(w) \sim \log n$.
The populations start at linkage equilibrium and, since all loci are unlinked,
we expect them to remain close to it.
The fitness breeding values will therefore be approximately log-normally distributed
within populations, with log fitness having variance ${\rm var}(\log w) = 2 \sum_k s^2 F_k (1 - F_k) \sim L s^2$, since each locus is an independent binomial variable.
Since $w$ is log-normal, the log of its variance is approximately 
$\log{\rm var}(w) \approx {\rm var}(\log w) + 2 \log \bar w$,
where $\bar w$ is its mean.
Near the replacement-level mean fitness of $\bar w \approx 2$ offspring, we therefore roughly need
${\rm var}(\log w) \sim \log n$ for the fit isogamous individuals to be hitting the 
limit of their reproductive capacity, or $L s^2 \sim \log n$.
We thus expect that the probability that anisogamy wins will depend primarily on
the relative magnitudes of $L s^2$ and $\log n$, i.e., on the compound parameter $L s^2 / \log n$.
This is confirmed by simulations: Fig.~\ref{fig:ls_sqr} shows that for fixed $n$ the 
probability of anisogamous fixation is a function of just $Ls^2$, while Fig.~\ref{fig:ls_sqr_n} shows that when $n$ is also varied the probability of
anisogamous fixation is a function of just the compound parameter $Ls^2/\log n$.
The above derivation was very rough, ignoring for instance the correction of 
$\approx 2 \log 2$, but as our model is a very simplified approximation of any real population, we are just focusing on finding the scaling relationships.

\section{Discussion}

In this work, we show that the faster spread of beneficial alleles
allowed by the presence of males can be sufficient 
to overcome their two-fold cost and maintain anisogamy in populations
adapting to environmental change. 
By comparing anisogamy to isogamy, we have carefully isolated the effect of
sex, rather than confounding it with the effects of recombination.
In our minimal model for sexual selection, the key parameter combination determining
whether anisogamy is favored over isogamy is the ratio of the variance
of potential fitness to the maximal reproductive capacity of isogamous individuals
and females, or, equivalently, $L s^2 / \log n$.
While we have cast our model in terms of male and female individuals,
it would apply equally well to hermaphroditic or monoecious species, or any other
mating system as long as each offspring receives resources primarily from one parent.

We have focused on the ability of a fully anisogamous population to outcompete
a similarly-sized isogamous population. 
To fully describe the role of rapid adaptation in the origin and maintenance of anisogamy, two extensions should be considered in the future. 
First, the assumption that anisogamous and isogamous populations start at the same
size should be relaxed.
It may be that the mechanism explored here allows the maintenance of anisogamy,
but does not allow it to spread when rare. 
Conversely, it may be that even when the effects discussed here are too weak to be effective under our 
starting conditions, they can still prevent isogamy from invading an anisogamous
population when the isogamous individuals are rare.
Second, instead of only considering competition between fully isogamous and fully
anisogamous populations, anisogamy should be treated as a quantitative trait, 
capable of evolving by degrees.
A model of competition between alleles that leads to only slightly different
degrees of anisogamy could lead to qualitatively different conditions,
as carriers of the different alleles would likely interbreed, potentially
breaking down the associations between the anisogamy locus and the directly-selected
loci \cite{kleiman2015}.

Our model is in some ways very similar to Trivers' original argument that 
the two-fold cost of males could be overcome if sexual selection were strong 
enough so that the average father would be much fitter
than the average mother \citep{trivers1976}.
However, his argument considered only one generation
and whether a female could maximize her number of successful daughters via
asexual or sexual reproduction, and therefore required potential fathers to be
at least twice as fit as mothers for sex to have an advantage.
Such a large gap is thought to be rare in natural populations \citep{kodricbrown1987}.
In our model, we show that because
the ``good genes'' contributed by fathers can compound over time, anisogamy can be 
favored over the course of multiple generations, even if over a single 
generation isogamy wins, e.g., in Fig.~\ref{fig:example_1run}, the anisogamous
population is initially increasing.
This means that the necessary gap between paternal and maternal fitness 
can be substantially smaller than in the Trivers' argument.

Our model does still require quite rapid adaptation for anisogamy to be favored:
we must have $L s^2 / \log n \ge \mathcal{O}(1)$, 
implying variance in fitness of at least $\mathcal{O}(1)$.
This is more than can be maintained at steady-state by the influx of beneficial
mutations \citep{weissman2012}, although it could be attained temporarily
during adaptation to large environmental shifts, as in our simulations. 
However, this raises the question of what happens when the environment is stable:
if anisogamy only has a temporary advantage, with isogamy having a two-fold 
advantage for long periods before and after, then this mechanism would not seem
to be able to contribute substantially to the maintenance of anisogamy.
There are two main ways in which this objection can fail. 

First, the environment
may always be shifting, either constantly exposing new formerly-neutral variation to selection, or fluctuating in direction of the same traits, as is observed on seasonal time scales in \textit{Droshophila} \citep{bergland2014}.
Selection could also be fluctuating over space, with anisogamous demes successfully 
adapting to environmental shifts while isogamous ones go extinct.
In this sense, our model can also be seen as falling in the general category of Red Queen models for the evolution of sex in continually adapting populations;
such models often consider evolving parasites as a source for the environmental shifts \citep{hamilton1990sexual}.
During each environmental shift, genetic hitchhiking will erode
the standing neutral variation that could potentially be exposed to selection
by future environments: roughly speaking, $N$ in Eq.~\eqref{standingvar} should be replaced by $N_e$, with $\log(N_e/N) \sim -L s^2$ \cite{weissman2012}. 
But this means that as long as $N \mu / n \gg 1$, i.e., the maternal mutation supply is large, there can still be plenty of genetic variation.

A second possibility is that the threshold rate of adaptation needed for this mechanism to maintain
anisogamy may be substantially lower than in our simulations.
We have focused on a minimal model of sexual selection, where on a gamete-by-gamete
level selection is the same in males and females, with the only difference arising because males have more gametes.
But sexual selection can involve much stronger selection on males even on a per-mating basis.
If we allow selection on males to be stronger by a factor $\alpha > 1$, 
this would reduce the threshold for anisogamy to win to $L s^2 / \log n \ge \mathcal{O}(1/\alpha^2)$, corresponding to a minimum rate of adaptation in females of 
only $\mathcal{O}(1/\alpha)$.
If we imagine, for example, that a mutation allowed dairy cows to reproduce 
isogamously with each other, this mutation would likely be disfavored,
because (artificial) sexual selection is so strong that effectively
all selection is taking place in bulls.
Note that this can be true even if the rate of increase in the population's mean milk production, i.e., the rate of adaptation, is not particularly high.
On the other hand, in natural populations, such stronger forms of sexual selection might also lead to genetic conflict,
in which the alleles that favor male mating success do not improve female fitness,
and might also involve direct costs to females in the selection process;
these effects would reduce the advantage provided by anisogamy.
The experimental evidence for how sexual selection interacts with 
adaptation to environmental shifts is mixed \cite{candolin2008}:
some studies have, indeed, found that it
accelerates natural selection (e.g., \cite{hollis2009}), 
while others argued that it impedes it (e.g., \cite{rundle2006,chenoweth2015}), or that it can do either
depending on the environment \cite{yun2017}.

The relevance of our model to the evolution of sex is ultimately an empirical question.
If our mechanism is, in fact, a significant contributor, we would expect that 
anisogamy would typically be more extreme in taxa that have undergone more rapid
adaptation in the past. However, this is a very difficult prediction to test. 
The rate of past adaptation is difficult to measure, particularly so for the
fitness flux \citep{mustonen2010}, the relevant quantity here,
as opposed to the rate of adaptive substitutions.
In addition, the rate of adaptation is likely to correlate with many other
factors, all of which could also select for or against anisogamy, confounding
the analysis. 
It may, therefore, make sense to begin by testing our mechanism's strength in
experimental populations.
Ideally, this would involve direct competition between individuals differing
only in their degree of anisogamy.
However, it may be experimentally more tractable 
to use closely related isogamous and anisogamous species, as are found in, for example,
the culturable filamentous fungus genus $\textit{Allomyces}$ \citep{phadke2017}.
By competing multiple isogamous and anisogamous species and strains against each
other under varying degrees of stress, one could test whether the advantage in 
adaptability conferred by anisogamy is large enough to consistently outweigh 
the idiosyncratic factors favoring one strain or another.
If so, this would be a powerful argument for the potential importance
of this mechanism in the evolution of anisogamous sex.

\section*{Acknowledgements}

CMH was supported by NIH grant 5 R01 EB022872 02 and the President's Fellowship at Princeton University. DBW was partially supported by a Mathematical Modeling of Living Systems Investigator Award from the Simons Foundation. IN was partially supported by the NSF grant PHY-1410978. IN and DBW acknowledge the hospitality of the Aspen Center for Physics, which is supported by the NSF grant PHY-1607611.

% Our model could be further extended by introducing non-random mating, removing discrete generational constraints, or introducing spatial inhomogeneities leading to compartmentalization. The latter may be able to explain co-existence of both mating modalities as is often observed, such as in the vegetative or sexual reproduction of plants. Here we stress that even the simplest model, accounting only for anisogamy, rather than for other phenomena usually correlated with it, is able to overcome the two fold cost of males, one of the oldest puzzles in evolutionary biology. 

%\section*{References}

\bibliography{mybibfile}

\end{document}